\documentclass{Interspeech}
\usepackage{multirow}

\interspeechcameraready 

\makeatletter
\renewcommand\anonaddress{%
  \ifinterspeechfinal
    {\centering
      \vskip-3pt
      \shortstack[c]{%
        \textsuperscript{1}Zhejiang University, China\quad
        \textsuperscript{2}UC Berkeley, United States\\
        \textsuperscript{3}Southern University of Science and Technology, China\quad
        \textsuperscript{4}UCSF, United States
      }\par
      \vskip-3pt
    }%
  \fi
}
\makeatother

\title{Seamless Dysfluent Speech Text Alignment for Disordered Speech Analysis}

\author[affiliation={ 1 }]{Zongli}{Ye}
\author[affiliation={ 2 }]{Jiachen}{Lian}
\author[affiliation={ 1 }]{Xuanru}{Zhou}
\author[affiliation={ 1 }]{Jinming}{Zhang}
\author[affiliation={ 3 }]{Haodong}{Li}
\author[affiliation={ 1 }]{Shuhe}{Li}
\author[affiliation={ 1 }]{Chenxu}{Guo}
\author[affiliation={ 2 }]{Anaisha}{Das}
\author[affiliation={ 2 }]{Peter}{Park}
\author[affiliation={ 4 }]{Zoe}{Ezzes}
\author[affiliation={ 4 }]{Jet}{Vonk}
\author[affiliation={ 4 }]{Brittany}{Morin}
\author[affiliation={ 4 }]{Rian}{Bogley}
\author[affiliation={ 4 }]{Lisa}{Wauters}
\author[affiliation={ 4 }]{Zachary}{Miller}
\author[affiliation={ 4 }]{Maria}{Gorno-Tempini}
\author[affiliation={ 2 }]{Gopala}{Anumanchipalli}

\affiliation{}{Zhejiang University}{China}
\affiliation{}{UC Berkeley}{United States}
\affiliation{}{UCSF}{United States}
\email{yezongli25@gmail.com, jiachenlian@berkeley.edu, gopala@berkeley.edu}

\keywords{speech pronunciation, dysfluency, forced alignment, clinical}

\usepackage{comment}
\usepackage[table,xcdraw]{xcolor}
\usepackage{hyperref}

\begin{document}

\maketitle

\begin{abstract} 
Accurate alignment of dysfluent speech with intended text is crucial for automating the diagnosis of neurodegenerative speech disorders. Traditional methods often fail to model phoneme similarities effectively, limiting their performance. In this work, we propose Neural LCS, a novel approach for dysfluent text-text and speech-text alignment. Neural LCS addresses key challenges, including partial alignment and context-aware similarity mapping, by leveraging robust phoneme-level modeling. We evaluate our method on a large-scale simulated dataset, generated using advanced data simulation techniques, and real PPA data. Neural LCS significantly outperforms state-of-the-art models in both alignment accuracy and dysfluent speech segmentation. Our results demonstrate the potential of Neural LCS to enhance automated systems for diagnosing and analyzing speech disorders, offering a more accurate and linguistically grounded solution for dysfluent speech alignment.
\end{abstract}

\section{Introduction}

The diagnosis and analysis of neurodegenerative speech disorders, such as primary progressive aphasia (PPA)~\cite{gorno2011classification-ppa}, traditionally depend on real-time MRIs (rtMRIs) and manual speech transcripts by speech-language pathologists (SLPs). Recent automated approaches for diagnosing and analyzing dysfluent speech have 
 focused on comparing uttered speech (actual spoken text) with lexical speech (intended text), with discrepancies termed dysfluencies like sound repetition, insertion, deletion, and substitution~\cite{lian2023unconstraineddysfluencymodelingdysfluent, lian2024hierarchicalspokenlanguagedysfluency}. Accurate identification of dysfluencies is essential for developing automated speech disorder diagnosis systems, relying on precise \textit{dysfluent speech-text alignment}.

Speech-text alignment, or forced alignment, maps speech tokens to corresponding text and identifies their temporal boundaries. It is crucial for tasks like text-to-speech (TTS) synthesis, data segmentation, phonetic research, and speech assessment. Various methods exist for forced alignment~\cite{graves2006connectionist-ctc, tian2022bayes-ctc, pratap2024scaling-mms, mcauliffe2017montreal-mfa, zhu2022phone-w2v2-alignment}. Typically, it assumes strong monotonicity and element-wise alignability, where each speech token is monotonically mapped to a text token based on similarity. While this works for fluent speech, \textit{aligning dysfluent or disordered speech requires different or stricter constraints}.

We illustrate an ideal dysfluent speech alignment and its role in enhancing automatic speech disorder diagnosis. Suppose a patient is instructed by an SLP to read "A pen on the table," with the phonetic ground truth: \texttt{/AH . P EH N . AA N . DH AH . T EY B AH L./}. A possible dysfluent transcription might be:
\texttt{/UH. UH. EY. P EN K N. AH N. DH AH. DH AH. T T T EY B AH L./}.
An ideal alignment would map actual pronunciations to intended phonemes as: \texttt{AH-(UH, UH, EY) . P-(P) EH-(EH K) N-(N) . AA-(AH) N-(N) . DH-(DH AH DH) AH-(AH) . T-(T, T, T) EY-(EY) B-(B) AH-(AH) L-(L)/}. This alignment process implicitly performs dysfluency detection. For instance, \texttt{EH-(EH K)} marks an insertion, while \texttt{AH-(UH, UH, EY)} reveals repetition of acoustically similar sounds. Unlike traditional fluent speech alignment, there are three challenges in dysfluent speech text alignment, which we outline next.

First, dysfluent speech is often only partially aligned with the text. For example, when pronouncing pen (/P EH N/), disordered speech might produce /P K EN N/, where the randomly inserted sound /K/ should not be aligned. Here, we represent speech using its transcribed phoneme sequence. Second, robust, context-aware similarity mapping is crucial, as exhaustive enumeration fails to address complexities like /B B P EN N/, where /P-B/ substitution indicates a voicing error, suggesting /B/ should align with /P/ based on articulatory, acoustic, or semantic similarity. However, accurate phonetic transcription remains challenging, as state-of-the-art phoneme recognition systems~\cite{wavlm-ctc, xu2021simple-w2v2-phoneme, li2020universal} struggle with atypical speech.

SSDM~\cite{ssdm, lian2024ssdm2.0} framed dysfluent speech alignment as a local sequence alignment problem, proposing the longest common subsequence (LCS) algorithm~\cite{hirschberg1977algorithms-lcs1} as a solution. Unlike global aligners like DTW~\cite{sakoe1971dynamic-dtw1}, which consider all tokens, LCS focuses on matching tokens, ignoring irrelevant ones. SSDM introduced the connectionist subsequence aligner (CSA) as a differentiable LCS, but preliminary results show minimal improvement over vanilla LCS, especially in phoneme similarity modeling, as CSA treats phonetically similar sounds as distinct. In this context, a robust, and linguistically grounded subsequence aligner is on the verge of emerging.

To provide experiments with larger-scale data that can more realistically reflect acoustic characteristics (such as phoneme pronunciation similarities), new data simulation methods have been explored. We inject disfluencies into VCTK~\cite{yamagishi2019vctk} text data based on similarity probabilities, generating a larger-scale disfluent dataset. Additionally, to ensure more natural text-speech data, we employ an LLM+TTS approach for synthesis.

In this work, we propose Neural LCS to handle the aforementioned problem. First, we focus on dysfluent text-to-text alignment. Our model outperforms significantly Hard LCS and DTW, especially at the phoneme level. Next, we introduce a speech-to-text alignment model based on Neural LCS and test it on dysfluent speech segmentation using simulated speech and PPA data. Our results surpass the current state-of-the-art models. To facilitate further research, we open-source our work at \href{https://github.com/Berkeley-Speech-Group/Neural-LCS.git}{\texttt{https://github.com/Berkeley-Speech-Group/N\\eural-LCS.git}}

\begin{figure*}[htbp]
    \centering
    \includegraphics[height=6.5cm, width=17cm]{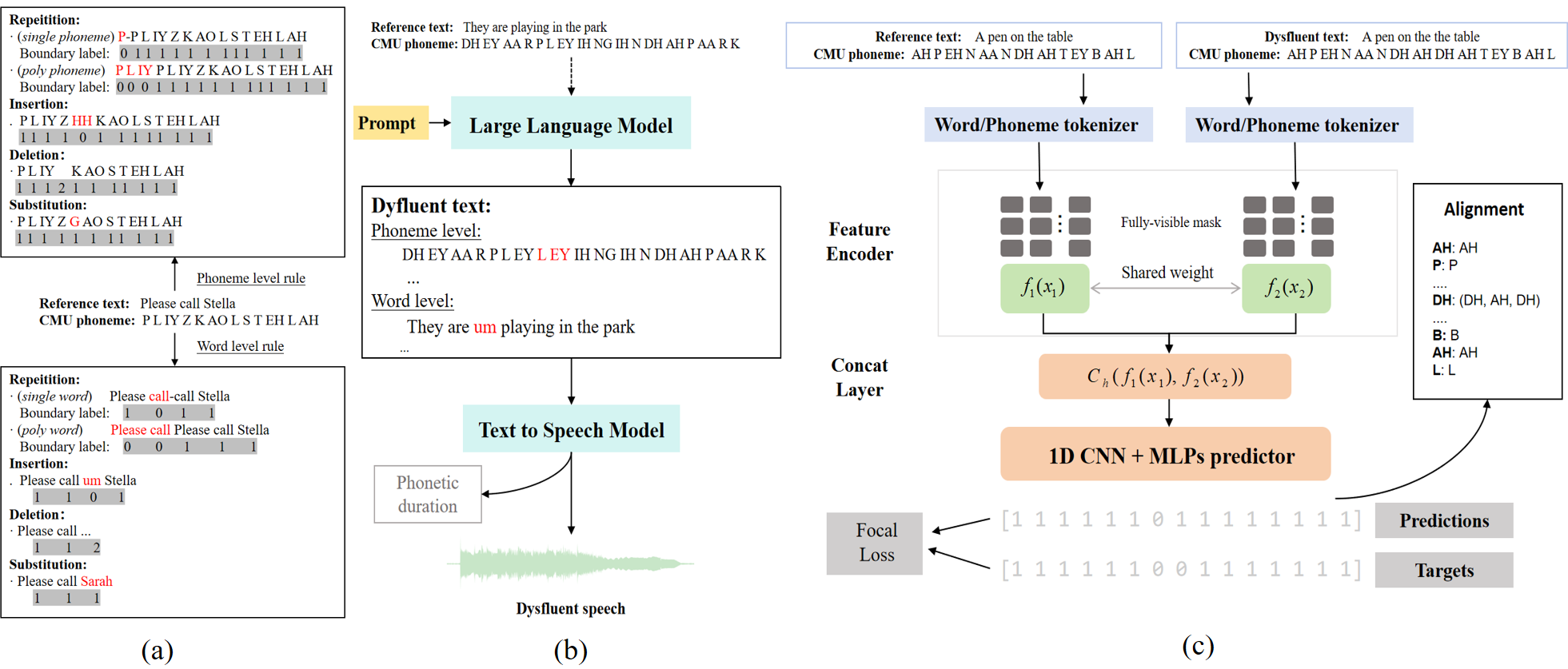}
    \caption{(a)shows our probabilistic random dysfluency injection and alignment label annotation. (b)shows how we combine LLMs and TTS to generate disfluent text and corresponding audio. Statistics of these two simulated data is shown in Table~\ref{table:dataset}.(c)illustrates the pipeline of Neural LCS model, 
the reference sequence and dysfluent sequence pass through the phoneme or word tokenizer, then pass through a siamese neural network, and finally output the alignment result sequence. }
    \label{fig:text-text lcs pipeline}
\end{figure*}

\section{Methods}
\subsection{Hard LCS}
Dysfluencies often have aligned targets, so applying the longest common subsequence (LCS) algorithm~\cite{hirschberg1977algorithms-lcs1} can automatically align dysfluent parts to corresponding phonemes or words. This works due to LCS's local alignment nature, as pointed out in~\cite{ssdm}. The standard LCS algorithm uses dynamic programming, requiring exact token matches, which ignores acoustic-phonetic similarity. We refer to this as \textbf{Hard LCS}.
\subsection{Dysfluency Simulation}
\subsubsection{Text-text data}
Our neural LCS algorithm is text-based, so we inject phonemic dysfluencies into text corpora to generate dysfluency data. We heuristically categorize CMU phonemes~\cite{cmudict} based on airflow patterns and articulatory mechanisms to capture phoneme similarities, as shown in Table~\ref{table:cat}. Phonemes in the same category tend to be more similar and often appear in phonetic errors.
\begin{table}[h!]
\centering
\renewcommand{\arraystretch}{1.}
\scriptsize
\begin{tabular}{@{}ll@{}}
\toprule
\textbf{Category}   & \textbf{CMU Phoneme}                                \\ \midrule
\textbf{Plosive}     & P, B, T, D, K, G             \\
\textbf{Fricative}   & F, V, TH, DH, S, Z, SH, ZH       \\
\textbf{Affricate}   & CH, JH                             \\
\textbf{Nasal}       & M, N, NG                     \\
\textbf{Liquid}      & L, R                                  \\
\textbf{Glide}       & W, Y                                    \\
\textbf{Vowel}       & AA, AE, AH, AO, AW, AY, EH, ER, EY, IH IY, OW, OY, UH, UW \\ \bottomrule
\end{tabular}
\caption{Phoneme Categories}
\label{table:cat}
\end{table}

\vspace{-18pt}

 Using VCTK text~\cite{yamagishi2019vctk} as a reference, we identify four types of dysfluencies that can be represented textually: \textit{Repetition}, \textit{Deletion}, \textit{Substitution}, \textit{Insertion}

\begin{figure*}[htbp]
    \centering
    \includegraphics[height=6.cm, width=14cm]{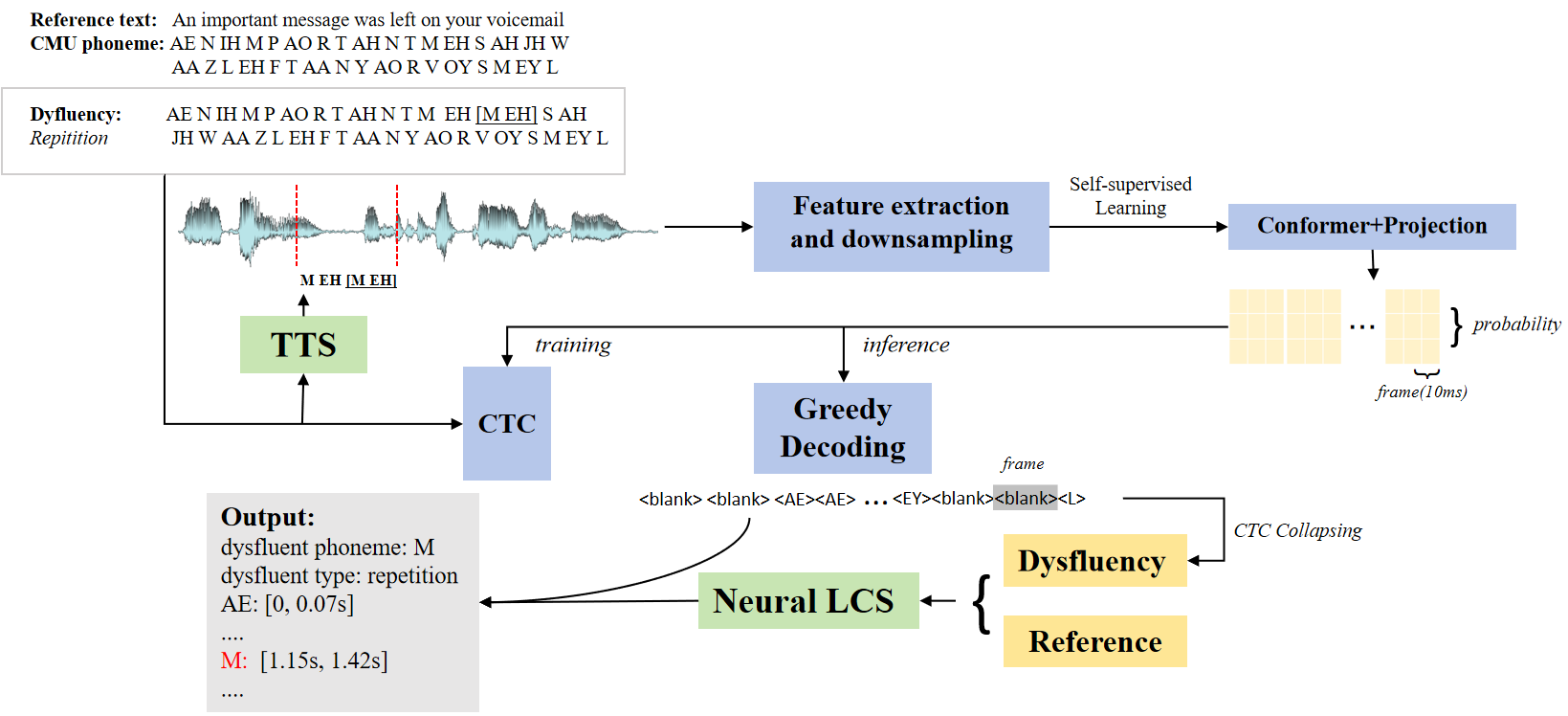}
    \caption{: Structure of Speech-text alignment model}
    \label{fig:text-speech lcs pipeline}
\end{figure*}

The word level simulation is similar, we replace letters or letter combinations in words according to phoneme similarity rules, and there are also four types of dysfluency.

We generate alignment labels between the original reference and dysfluent sequences. We use 1 to mark aligned boundaries in the reference, 0 for dysfluent units within the aligned part, and 2 for missing phonemes or words. The alignment label sequence includes these three labels. The text-text data simulation process is shown in Fig.~\ref{fig:text-text lcs pipeline}(a).

\vspace{-5pt}

\subsubsection{Text-speech data}

Probability-based random injection of dysfluent parts into text may not be natural enough for the actual generated audio~\cite{zhou2024yolostutterendtoendregionwisespeech, zhou2024stutter, zhou2024timetokensbenchmarkingendtoend, zhang2025analysisevaluationsyntheticdata}. As large language models (LLMs) have demonstrated exceptional capabilities in natural language generation. In this part, we explored using LLMs to help us generate more natural and coherent dysfluent speech text. we leverage Claude~\cite{anthropic2024claude} to simulate dysfluencies in text. By guiding the model to introduce various forms of dysfluency mentioned in Sec.2.2.1  we can generate dysfluent text data that closely resembles real-world speech disorders. We then use VITS~\cite{kim2021conditionalvariationalautoencoderadversarial} as TTS model to generate a new dysfluency audio dataset, we call it \textbf{LLM disorder}. In the TTS process, we obtain the vector representing the number of time frames occupied by each phoneme in the audio. By mapping the IPA phonemes to their corresponding CMU phonemes, we derive the true alignment between the audio time and its reference phonemes.

\begin{table}[htbp]
\centering
\setlength{\tabcolsep}{3.8pt} 
\renewcommand{\arraystretch}{1.1}
\scriptsize
\begin{tabular}{@{}ccc@{}}
\toprule
\textbf{Simulated Dataset}        & \textbf{Total Amount of Text} & \textbf{Total Hours of Audio} \\ \midrule
\textbf{Text-text data} & 110.7\(\times\)10\(^{4}\) sentences       & -                    \\
\textbf{LLM disorder} & 10\(^4\)\(\times\)50 (50 speakers) sentences          & 1152.86h              \\ \bottomrule
\end{tabular}
\caption{Statistics of Simulated Datasets}
\label{table:dataset}
\end{table}

\vspace{-15pt}

\subsection{Neural LCS}
To address the alignment inaccuracies in Hard LCS due to neglecting pronunciation similarity, we designed \textbf{Neural LCS}. It takes dual sequence inputs and outputs alignment labels (0, 1, or 2, as in Sec. 2.2.1). Using a siamese network framework~\cite{bromley1994siamese, 9893751}, the input token sequences are processed through a shared Feature encoder. The resulting sequence representations are concatenated and passed through a 1D CNN and MLPs with fully connected layers and ReLU activation. A softmax function is applied to the output. The entire pipeline is shown in Fig.~\ref{fig:text-text lcs pipeline}(c), with further details discussed below.
\subsubsection{Feature Encoder}
We use the default T5~\cite{raffel2023exploringlimitstransferlearning} feature encoder with a fully-visible mask, enabling all tokens to attend to each other. For phoneme-level tokenization, we implemented a custom tokenizer based on the CMU phoneme dictionary, while for word-level tokenization, we used the default T5-small tokenizer.
\vspace{-5pt}
\subsubsection{Training Objective}
Our alignment label exhibits an imbalance in class distribution. 
If a text contains less dysfluency, label 1 accounts for most of its corresponding label sequence. Thus we employ Focal Loss~\cite{lin2018focallossdenseobject} $\mathcal{L}_{\text{FL}}$, which is defined as:
\begin{equation}
\mathcal{L}_{\text{FL}}(p_t) = -\alpha (1 - p_t)^\gamma \log(p_t)
\label{eq:focal_loss}
\end{equation}
where \( p_t \) represents the predicted probability of the true class, \( \alpha \) is a weighting factor for class balance, and \( \gamma \) controls the down-weighting of well-classified samples. By focusing more on hard-to-classify instances, Focal Loss mitigates the impact of class imbalance and enhances the model's ability to learn from underrepresented classes. In this work, we let \([\alpha_0, \alpha_1, \alpha_2] = [0.5, 0.1, 0.8]\), \(\gamma = 3\).
\vspace{-5pt}
\subsection{Speech-text alignment}
In Sec.2.2.2, we obtained text-speech data generated by LLM. In this part, we propose \textbf{STA}(Speech-text alignment) model. we use Neural LCS as a basic component to implement an end-to-end framework for aligning audio and text, which can directly segment and detect dysfluencies from the speech signal. The entire paradigm is shown in Fig.~\ref{fig:text-speech lcs pipeline}. 

For phoneme level process, wax2vec2.0 feature extractor ~\cite{schneider2019wav2vecunsupervisedpretrainingspeech} accepts the speech signal generated by LLM and TTS model, it converts the feature dimension of each frame of signal to the length of CMU dictionary + 1 via basic conformer and projection layer, and is trained with the source dysfluent sequence of the generated speech through CTC-Loss~\cite{graves2006connectionist-ctc}. 

During inference, we apply greedy decoding on CTC emission matrix. This gives us the alignment between the dysfluent sequence and the audio timeline. Next, we align the collapsed dysfluent sequence with the reference text, which provides the alignment between the reference and the audio timeline. This allows us to accurately segment the audio based on the reference and speech.

The advantage of using Neural LCS as the aligner is that when the dysfluent unit transcription is inaccurate, the reference and speech can still be matched through soft alignment to achieve more accurate segmentation.

\definecolor{brightturquoise}{rgb}{0.8, 0.9, 1.0}

\begin{figure*}[htbp]
    \centering
    \includegraphics[height=3.5cm, width=17cm]{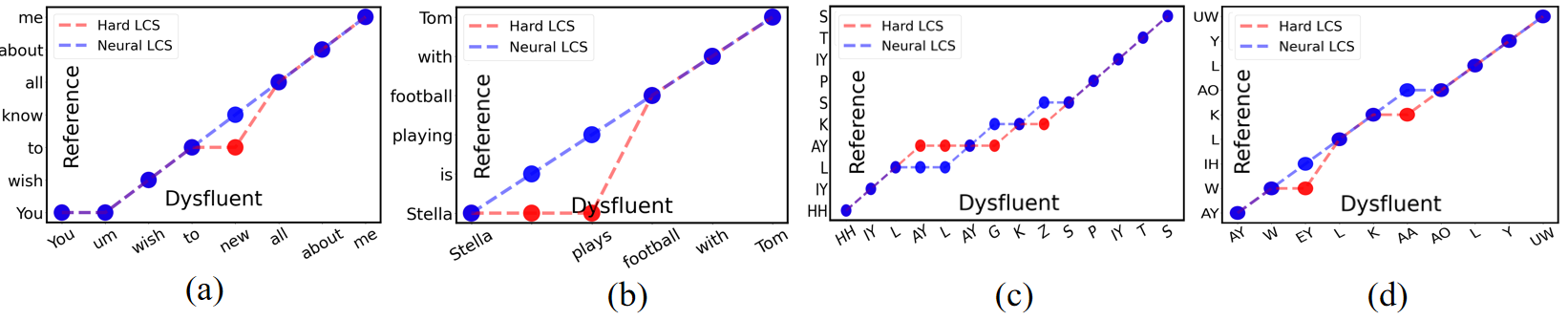}
    \caption{: (a) shows that our word-level Neural LCS model captures acoustic similarities between words, even if they contain different letters, like swiftly-wishy. (b) demonstrates the model's ability to capture morphological similarities, even without similar pronunciations, like plays-playing. (c) highlights the phoneme-level model's ability to capture consonant similarities, such as K-G, S-Z. (d) shows that the model captures vowel similarities, like IH-EY, AO-AA.}
    \label{fig:nnlcs result}
\end{figure*}

\section{Experiments}

\subsection{Dataset}
\textbf{(1) VCTK~\cite{yamagishi2019vctk}: }it includes 109 native English speakers with accented speech. It's text is used in our text-text data simulation as mentioned in Sec.2.2.1.\textbf{(2) LLM disorder: }We use LLM+TTS to generate large scale more natural dysfluent text-speech data. The detail is shown in Sec.2.2.2.\textbf{(3) PPA Speech~\cite{gorno2011classification-ppa}: }it includes recordings from 38 participants diagnosed with Primary Progressive Aphasia (PPA). Participants were asked to read the "grandfather passage," resulting in approximately one hour of speech in total.

\vspace{-5pt}

\subsection{Training Details}
We performed a randomized 90/10 train/test split on both text-text and text-speech data. The Neural LCS model was trained with a batch size of 32 for 15 epochs(both phoneme level and word level), totaling 16 hours on an RTX A6000, using Adam optimization with a learning rate of 1e-4, without dropout or weight decay. The phoneme-level speech-text alignment model, with a batch size of 1, was trained for 10 epochs, totaling 50 hours on the same GPU, under the same optimization settings.

\vspace{-5pt}

\subsection{Results}
\subsubsection{Speech-text alignment}
Comparison diagram between Neural LCS algorithm and Hard LCS is shown in Fig.~\ref{fig:nnlcs result}.It can be seen that our model performs well in handling the alignment of similar phonemes or words. For the phoneme level, the model can more accurately capture the pronunciation similarity between vowels and the pronunciation similarity between consonants. For the word level, the model can focus on the similarity of pronunciation or letter composition between words to achieve a more reasonable alignment. Compared with the Hard LCS algorithm, the Neural LCS model can better combine the related features of phonemes or words to achieve alignment. In addition, in the example shown in Fig.~\ref{fig:nnlcs result}, our model can also accurately identify various dysfluency types(talked in Sec. 2.2.1) through soft alignment.

We apply DTW, Hard LCS, and Neural LCS to the test set of text-text data and LLM-generated text to compare their alignment accuracy with the reference text. Results are shown in Table~\ref{table:alg cmp}. On both test sets, Neural LCS significantly improves alignment of dysfluent phonemes and words over traditional DTW and Hard LCS, consistent with our analysis. It also detects dysfluency types more accurately. Additionally, Neural LCS performs better at the phoneme level, likely due to the added complexity of word-level alignment, which involves both morphological and acoustic features.

\begin{table}[htbp]
\centering
\setlength{\tabcolsep}{8.8pt} 
\renewcommand{\arraystretch}{1.2} 
\footnotesize
\begin{tabular}{cccc}
\hline
\textbf{Level}                    & \textbf{Method}  & \textbf{text-text data}                  & \textbf{LLM text}                        \\ \hline
\multirow{3}{*}{Phoneme} & DTW        & 33.47\%                           & 54.80\%                           \\ 
                         & Hard LCS   & 24.78\%                           & 43.53\%                           \\ 
                         & Neural LCS & \textbf{72.55\%} & \textbf{90.96\%} \\ \hline
\multirow{3}{*}{Word}    & DTW        & 58.65\%                           & 62.42\%                           \\  
                         & Hard LCS   & 58.47\%                           & 60.67\%                           \\ 
                         & Neural LCS & \textbf{68.44\%} & \textbf{75.07\%} \\ \hline
\end{tabular}

\caption{Comparison of Different Methods}

\label{table:alg cmp}
\end{table}

\vspace{-25pt}

\subsubsection{Speech-text alignment}
We conduct our STA (Speech-text Alignment) model inference on our proposed LLM disorder data and PPA Speech, using YOLO-Stutter~\cite{zhou2024yolostutterendtoendregionwisespeech}, an open-sourced state-of-the-art model for dysfluency boundary detection in speech, as the baseline. We evaluate the models using \textbf{Boundary Loss}(BL): the mean squared error between the predicted and actual boundaries of the dysfluent regions.

\begin{table}[h]
    \centering
    \scriptsize
    \setlength{\tabcolsep}{5.8pt} 
    \renewcommand{\arraystretch}{1.1} 
    \begin{tabular}{l l|c|c|c|c} 
     \toprule
      \textbf{Methods}& \textbf{Evaluated Dataset} & \multicolumn{1}{c|}{\textbf{Rep}}& \multicolumn{1}{c|}{\textbf{Del}} & \multicolumn{1}{c|}{\textbf{Sub}}& \multicolumn{1}{c}{\textbf{Ins}}\\
      \rowcolor{brightturquoise}
    & & BL& BL & BL  & BL   \\
    \hline
    \hline
    YOLO-Stutter & LLM disorder&27ms & \textbf{13ms} & 10ms & 50ms \\
    STA model & LLM disorder& \textbf{10ms} & 27ms & \textbf{8ms} & \textbf{23ms} \\
    \midrule
    YOLO-Stutter & PPA Speech & \multicolumn{4}{c}{21ms} \\
    STA model & PPA Speech & \multicolumn{4}{c}{\textbf{17ms}} \\
    \bottomrule
    \end{tabular}
    \caption{Boundary Loss(BL) of the four dysfluency types}
    \label{evaluate-comparison}
\end{table}
\vspace{-12pt}

As indicated in Table~\ref{evaluate-comparison}, except deletion detection, our STA model outperforms YOLO-Stutter in terms of the BL metric. In particular, there have been significant improvements in repetition and Insertion, which means that our STA model can more accurately match the dysfluent parts of speech with timestamps. Notably, our STA model adopts full sequence alignment, so except for the unfluent parts, we can perform speech alignment on all phonemes in the reference.

\subsection{Ablation experiments}
To investigate the impact of the proportions of different dysfluency types quantities on training results, we selected four
different proportions except for average on our text-text data, as follows: P = [Repetition, Insertion, Deletion, Substitution], P1=[1:1.5:1:1.5], P2= [1:1.5:1.5:1], P3=[1:1:1.5:1.5], P4=[1:1:1.2:1]. Table~\ref{table:type-acc} shows the type-specific accuracy on LLM disorder text (Mix refers to multiple dysfluency types in a sentence). Despite proportion adjustments, repetition accuracy remained high and stable, while substitution stayed relatively low. Pairwise comparison reveals that increasing the substitution proportion improves its accuracy but lowers insertion accuracy, and vice versa. Increasing the deletion proportion has a minimal impact on other types.

\begin{table}[htbp]
    \centering
    \setlength{\tabcolsep}{6.3pt}
    \renewcommand{\arraystretch}{0.9} 
    \scriptsize
    \begin{tabular}{lccccc}
        \toprule
        \textbf{Proportions} & \textbf{Rep} & \textbf{Ins} & \textbf{Del} & \textbf{Sub} & \textbf{Mix} \\
        \midrule
        Average & 96.23\% & 83.85\% & 93.42\% & 91.25\% & 90.96\%\\
        $P_1$ & 96.40\% & 81.95\% & 92.09\% & 92.20\% & 90.64\% \\
        $P_2$ & 95.79\% & 83.93\% & 93.74\% & 86.61\% & 89.68\%\\
        $P_3$ & 96.10\% & 81.37\% & 95.39\% & 93.22\% & 91.56\%\\
        $P_4$ & 96.51\% & 82.55\% & 95.04\% & 92.77\% & 90.14\%\\
        \bottomrule
    \end{tabular}
    \caption{Type-specific accuracy of different proportions}
    \label{table:type-acc}
\end{table}
\vspace{-25pt}


\section{Conclusion and Future Work}
We propose Neural LCS, a novel method for aligning dysfluent speech that overcomes limitations of existing approaches. It operates in two modes: (1) aligning transcribed phonemes or words to reference sequences using acoustic or morphological cues, and (2) directly segmenting speech based on the reference. By leveraging acoustic-phonetic similarity, Neural LCS significantly improves dysfluency detection and segmentation accuracy. We contribute two simulated dysfluent corpora: \textit{text-text data} and \textit{LLM Disorder}, containing acoustic features and naturally disfluent text/audio with high-quality annotations. Experiments on both simulated and real disordered speech datasets demonstrate Neural LCS outperforms existing models in speech-text alignment, making it valuable for clinical and research applications. Neural LCS in essence tackles the word and phoneme allophony issue~\cite{boomershine2008impact-allophony, choi2025leveragingallophonyselfsupervisedspeech}, and it would be helpful to also explore better decoder~\cite{guo2025dysfluentwfstframeworkzeroshot} or
phoneme similarity models either in kinematics systems~\cite{cho2024jstsp, wu23k_interspeech} or gestural systems~\cite{ssdm, lian22bcsnmf, lian2023factor}.
\section{Acknowledgements}
Thanks for support from UC Noyce Initiative, Society of Hellman Fellows, NIH/NIDCD, and the Schwab Innovation fund.


\newpage
\bibliographystyle{IEEEtran}
\bibliography{mybib}

\begin{thebibliography}{10}
\providecommand{\url}[1]{#1}
\csname url@samestyle\endcsname
\providecommand{\newblock}{\relax}
\providecommand{\bibinfo}[2]{#2}
\providecommand{\BIBentrySTDinterwordspacing}{\spaceskip=0pt\relax}
\providecommand{\BIBentryALTinterwordstretchfactor}{4}
\providecommand{\BIBentryALTinterwordspacing}{\spaceskip=\fontdimen2\font plus
\BIBentryALTinterwordstretchfactor\fontdimen3\font minus \fontdimen4\font\relax}
\providecommand{\BIBforeignlanguage}[2]{{%
\expandafter\ifx\csname l@#1\endcsname\relax
\typeout{** WARNING: IEEEtran.bst: No hyphenation pattern has been}%
\typeout{** loaded for the language `#1'. Using the pattern for}%
\typeout{** the default language instead.}%
\else
\language=\csname l@#1\endcsname
\fi
#2}}
\providecommand{\BIBdecl}{\relax}
\BIBdecl

\bibitem{lian2023unconstraineddysfluencymodelingdysfluent}
J.~Lian, C.~Feng, N.~Farooqi, S.~Li, A.~Kashyap, C.~J. Cho, P.~Wu, R.~Netzorg, T.~Li, and G.~K. Anumanchipalli, ``Unconstrained dysfluency modeling for dysfluent speech transcription and detection,'' in \emph{2023 IEEE Automatic Speech Recognition and Understanding Workshop (ASRU)}, 2023, pp. 1--8.

\bibitem{lian2024hierarchicalspokenlanguagedysfluency}
J.~Lian and G.~Anumanchipalli, ``Towards hierarchical spoken language disfluency modeling,'' in \emph{Proceedings of the 18th Conference of the European Chapter of the Association for Computational Linguistics}, 2024, pp. 539--551.

\bibitem{6349744}
J.~Pálfy and J.~Pospíchal, ``Pattern search in dysfluent speech,'' in \emph{2012 IEEE International Workshop on Machine Learning for Signal Processing}, 2012, pp. 1--6.

\bibitem{kouzelis2023weaklysupervisedforcedalignmentdisfluent}
T.~Kouzelis, G.~Paraskevopoulos, A.~Katsamanis, and V.~Katsouros, ``Weakly-supervised forced alignment of disfluent speech using phoneme-level modeling,'' in \emph{Interspeech}, 2023.

\bibitem{ssdm}
J.~Lian, X.~Zhou, Z.~Ezzes, J.~Vonk, B.~Morin, D.~P. Baquirin, Z.~Miller, M.~L. Gorno~Tempini, and G.~Anumanchipalli, ``Ssdm: Scalable speech dysfluency modeling,'' in \emph{Advances in Neural Information Processing Systems}, vol.~37, 2024.

\bibitem{lian2024ssdm2.0}
J.~Lian, X.~Zhou, Z.~Ezzes, J.~Vonk, B.~Morin, D.~Baquirin, Z.~Mille, M.~L.~G. Tempini, and G.~K. Anumanchipalli, ``Ssdm 2.0: Time-accurate speech rich transcription with non-fluencies,'' \emph{arXiv preprint arXiv:2412.00265}, 2024.

\bibitem{wiki:lcs}
\BIBentryALTinterwordspacing
Wikipedia, ``Longest common subsequence.'' [Online]. Available: \url{https://en.wikipedia.org/wiki/Longest_common_subsequence}
\BIBentrySTDinterwordspacing

\bibitem{zhou2024yolostutterendtoendregionwisespeech}
X.~Zhou, A.~Kashyap, S.~Li, A.~Sharma, B.~Morin, D.~Baquirin, J.~Vonk, Z.~Ezzes, Z.~Miller, M.~Tempini, J.~Lian, and G.~Anumanchipalli, ``Yolo-stutter: End-to-end region-wise speech dysfluency detection,'' in \emph{Interspeech 2024}, 2024, pp. 937--941.

\bibitem{zhou2024stutter}
X.~Zhou, C.~J. Cho, A.~Sharma, B.~Morin, D.~Baquirin, J.~Vonk, Z.~Ezzes, Z.~Miller, B.~L. Tee, M.~L. Gorno-Tempini \emph{et~al.}, ``Stutter-solver: End-to-end multi-lingual dysfluency detection,'' in \emph{2024 IEEE Spoken Language Technology Workshop (SLT)}.\hskip 1em plus 0.5em minus 0.4em\relax IEEE, 2024, pp. 1039--1046.

\bibitem{mcauliffe2017montreal-mfa}
M.~McAuliffe, M.~Socolof, S.~Mihuc, M.~Wagner, and M.~Sonderegger, ``Montreal forced aligner: Trainable text-speech alignment using kaldi.'' in \emph{Interspeech}, vol. 2017, 2017, pp. 498--502.

\bibitem{zhou2024timetokensbenchmarkingendtoend}
\BIBentryALTinterwordspacing
X.~Zhou, J.~Lian, C.~J. Cho, J.~Liu, Z.~Ye, J.~Zhang, B.~Morin, D.~Baquirin, J.~Vonk, Z.~Ezzes, Z.~Miller, M.~L.~G. Tempini, and G.~Anumanchipalli, ``Time and tokens: Benchmarking end-to-end speech dysfluency detection,'' 2024. [Online]. Available: \url{https://arxiv.org/abs/2409.13582}
\BIBentrySTDinterwordspacing

\bibitem{lian2023factor}
J.~Lian, A.~W. Black, Y.~Lu, L.~Goldstein, S.~Watanabe, and G.~K. Anumanchipalli, ``Articulatory representation learning via joint factor analysis and neural matrix factorization,'' in \emph{ICASSP 2023-2023 IEEE International Conference on Acoustics, Speech and Signal Processing (ICASSP)}.\hskip 1em plus 0.5em minus 0.4em\relax IEEE, 2023, pp. 1--5.

\bibitem{lian22bcsnmf}
J.~Lian, A.~W. Black, L.~Goldstein, and G.~K. Anumanchipalli, ``{Deep Neural Convolutive Matrix Factorization for Articulatory Representation Decomposition},'' in \emph{Proc. Interspeech 2022}, 2022, pp. 4686--4690.

\bibitem{wu23k_interspeech}
P.~Wu, T.~Li, Y.~Lu, Y.~Zhang, J.~Lian, A.~W. Black, L.~Goldstein, S.~Watanabe, and G.~K. Anumanchipalli, ``{Deep Speech Synthesis from MRI-Based Articulatory Representations},'' in \emph{Proc. INTERSPEECH 2023}, 2023, pp. 5132--5136.

\bibitem{barrault2023seamlessm4t}
L.~Barrault, Y.-A. Chung, M.~C. Meglioli, D.~Dale, N.~Dong, P.-A. Duquenne, H.~Elsahar, H.~Gong, K.~Heffernan, J.~Hoffman \emph{et~al.}, ``Seamlessm4t-massively multilingual \& multimodal machine translation,'' \emph{arXiv preprint arXiv:2308.11596}, 2023.

\bibitem{graves2006connectionist-ctc}
A.~Graves, S.~Fern{\'a}ndez, F.~Gomez, and J.~Schmidhuber, ``Connectionist temporal classification: labelling unsegmented sequence data with recurrent neural networks,'' in \emph{Proceedings of the 23rd international conference on Machine learning}, 2006, pp. 369--376.

\bibitem{tian2022bayes-ctc}
J.~Tian, B.~Yan, J.~Yu, C.~Weng, D.~Yu, and S.~Watanabe, ``Bayes risk ctc: Controllable ctc alignment in sequence-to-sequence tasks,'' \emph{ICLR}, 2022.

\bibitem{cho2024jstsp}
C.~J. Cho, P.~Wu, T.~S. Prabhune, D.~Agarwal, and G.~K. Anumanchipalli, ``Coding speech through vocal tract kinematics,'' in \emph{IEEE JSTSP}, 2025.

\bibitem{yamagishi2019vctk}
J.~Yamagishi, C.~Veaux, and K.~MacDonald, ``Cstr vctk corpus: English multi-speaker corpus for cstr voice cloning toolkit (version 0.92),'' 2019, [sound], University of Edinburgh, The Centre for Speech Technology Research (CSTR).

\bibitem{zhu2022phone-w2v2-alignment}
J.~Zhu, C.~Zhang, and D.~Jurgens, ``Phone-to-audio alignment without text: A semi-supervised approach,'' in \emph{ICASSP 2022-2022 IEEE International Conference on Acoustics, Speech and Signal Processing (ICASSP)}.\hskip 1em plus 0.5em minus 0.4em\relax IEEE, 2022, pp. 8167--8171.

\bibitem{pratap2024scaling-mms}
V.~Pratap, A.~Tjandra, B.~Shi, P.~Tomasello, A.~Babu, S.~Kundu, A.~Elkahky, Z.~Ni, A.~Vyas, M.~Fazel-Zarandi \emph{et~al.}, ``Scaling speech technology to 1,000+ languages,'' \emph{Journal of Machine Learning Research}, vol.~25, no.~97, pp. 1--52, 2024.

\bibitem{gorno2011classification-ppa}
M.~L. Gorno-Tempini, A.~E. Hillis, S.~Weintraub, A.~Kertesz, M.~Mendez, S.~F. Cappa, J.~M. Ogar, J.~D. Rohrer, S.~Black, B.~F. Boeve \emph{et~al.}, ``Classification of primary progressive aphasia and its variants,'' \emph{Neurology}, vol.~76, no.~11, pp. 1006--1014, 2011.

\bibitem{wavlm-ctc}
``Wavlm-ctc-hugginface,'' \url{https://huggingface.co/microsoft/wavlm-large}.

\bibitem{xu2021simple-w2v2-phoneme}
Q.~Xu, A.~Baevski, and M.~Auli, ``Simple and effective zero-shot cross-lingual phoneme recognition,'' \emph{Interspeech}, 2022.

\bibitem{li2020universal}
X.~Li, S.~Dalmia, J.~Li, M.~Lee, P.~Littell, J.~Yao, A.~Anastasopoulos, D.~R. Mortensen, G.~Neubig, A.~W. Black \emph{et~al.}, ``Universal phone recognition with a multilingual allophone system,'' in \emph{ICASSP}.\hskip 1em plus 0.5em minus 0.4em\relax IEEE, 2020, pp. 8249--8253.

\bibitem{hirschberg1977algorithms-lcs1}
D.~S. Hirschberg, ``Algorithms for the longest common subsequence problem,'' \emph{Journal of the ACM (JACM)}, vol.~24, no.~4, pp. 664--675, 1977.

\bibitem{sakoe1971dynamic-dtw1}
H.~Sakoe, ``Dynamic-programming approach to continuous speech recognition,'' in \emph{1971 Proc. the International Congress of Acoustics, Budapest}, 1971.

\bibitem{cmudict}
\BIBentryALTinterwordspacing
C.~M. University, ``Cmu phoneme dictionary.'' [Online]. Available: \url{http://www.speech.cs.cmu.edu/cgi-bin/cmudict}
\BIBentrySTDinterwordspacing

\bibitem{anthropic2024claude}
\BIBentryALTinterwordspacing
Anthropic, ``Model card addendum: Claude 3.5 haiku and upgraded claude 3.5 sonnet,'' 2024. [Online]. Available: \url{https://www.anthropic.com}
\BIBentrySTDinterwordspacing

\bibitem{kim2021conditionalvariationalautoencoderadversarial}
J.~Kim, J.~Kong, and J.~Son, ``Conditional variational autoencoder with adversarial learning for end-to-end text-to-speech,'' \emph{International Conference on Machine learning}, 2021.

\bibitem{bromley1994siamese}
J.~Bromley, I.~Guyon, Y.~LeCun, E.~S{\"a}ckinger, and R.~Shah, ``Signature verification using a {Siamese} time delay neural network,'' in \emph{Advances in Neural Information Processing Systems (NeurIPS)}, vol.~6, 1994, pp. 737--744.

\bibitem{9893751}
Y.~Li, C.~L.~P. Chen, and T.~Zhang, ``A survey on siamese network: Methodologies, applications, and opportunities,'' \emph{IEEE Transactions on Artificial Intelligence}, vol.~3, no.~6, pp. 994--1014, 2022.

\bibitem{raffel2023exploringlimitstransferlearning}
\BIBentryALTinterwordspacing
C.~Raffel, N.~Shazeer, A.~Roberts, K.~Lee, S.~Narang, M.~Matena, Y.~Zhou, W.~Li, and P.~J. Liu, ``Exploring the limits of transfer learning with a unified text-to-text transformer,'' 2023. [Online]. Available: \url{https://arxiv.org/abs/1910.10683}
\BIBentrySTDinterwordspacing

\bibitem{lin2018focallossdenseobject}
\BIBentryALTinterwordspacing
T.-Y. Lin, P.~Goyal, R.~Girshick, K.~He, and P.~Dollár, ``Focal loss for dense object detection,'' 2018. [Online]. Available: \url{https://arxiv.org/abs/1708.02002}
\BIBentrySTDinterwordspacing

\bibitem{schneider2019wav2vecunsupervisedpretrainingspeech}
\BIBentryALTinterwordspacing
S.~Schneider, A.~Baevski, R.~Collobert, and M.~Auli, ``wav2vec: Unsupervised pre-training for speech recognition,'' 2019. [Online]. Available: \url{https://arxiv.org/abs/1904.05862}
\BIBentrySTDinterwordspacing

\bibitem{boomershine2008impact-allophony}
A.~Boomershine, K.~C. Hall, E.~Hume, and K.~Johnson, ``The impact of allophony versus contrast on speech perception,'' \emph{Contrast in phonology}, pp. 143--172, 2008.

\bibitem{choi2025leveragingallophonyselfsupervisedspeech}
K.~Choi, E.~Yeo, K.~Chang, S.~Watanabe, and D.~Mortensen, ``Leveraging allophony in self-supervised speech models for atypical pronunciation assessment,'' in \emph{NAACL}, 2025.

\bibitem{zhang2025analysisevaluationsyntheticdata}
\BIBentryALTinterwordspacing
J.~Zhang, X.~Zhou, J.~Lian, S.~Li, W.~Li, Z.~Ezzes, R.~Bogley, L.~Wauters, Z.~Miller, J.~Vonk, B.~Morin, M.~Gorno-Tempini, and G.~Anumanchipalli, ``Analysis and evaluation of synthetic data generation in speech dysfluency detection,'' 2025. [Online]. Available: \url{https://arxiv.org/abs/2505.22029}
\BIBentrySTDinterwordspacing

\bibitem{guo2025dysfluentwfstframeworkzeroshot}
\BIBentryALTinterwordspacing
C.~Guo, J.~Lian, X.~Zhou, J.~Zhang, S.~Li, Z.~Ye, H.~J. Park, A.~Das, Z.~Ezzes, J.~Vonk, B.~Morin, R.~Bogley, L.~Wauters, Z.~Miller, M.~Gorno-Tempini, and G.~Anumanchipalli, ``Dysfluent wfst: A framework for zero-shot speech dysfluency transcription and detection,'' 2025. [Online]. Available: \url{https://arxiv.org/abs/2505.16351}
\BIBentrySTDinterwordspacing

\end{thebibliography}

\end{document}